\documentclass[twocolumn,showpacs,aps,prd,floatfix]{revtex4}

\usepackage{graphicx}
\usepackage{dcolumn}
\usepackage{amsmath}
\usepackage{epsfig}

\RequirePackage{xspace}

\usepackage{relsize}
\def\babar{\mbox{\slshape B\kern-0.1em{\smaller A}\kern-0.1em
    B\kern-0.1em{\smaller A\kern-0.2em R}}}

\def\epem       {\ensuremath{e^+e^-}\xspace}

\def\qqbar {\ensuremath{q\overline q}\xspace}

\def\piz   {\ensuremath{\pi^0}\xspace}

\def\pip   {\ensuremath{\pi^+}\xspace}
\def\pim   {\ensuremath{\pi^-}\xspace}

\def\Kbar  {\kern 0.2em\overline{\kern -0.2em K}{}\xspace}

\def\Kz    {\ensuremath{K^0}\xspace}
\def\Kzb   {\ensuremath{\Kbar^0}\xspace}
\def\KzKzb {\ensuremath{\Kz \kern -0.16em \Kzb}\xspace}
\def\Kp    {\ensuremath{K^+}\xspace}
\def\Km    {\ensuremath{K^-}\xspace}

\def\KpKm  {\ensuremath{\Kp \kern -0.16em \Km}\xspace}
\def\KS    {\ensuremath{K^0_{\scriptscriptstyle S}}\xspace}

\def\Kstar   {\ensuremath{K^*}\xspace}

\def\Kstarm  {\ensuremath{K^{*-}}\xspace}
\def\Kstarpm {\ensuremath{K^{*\pm}}\xspace}

\def\Dbar    {\kern 0.2em\overline{\kern -0.2em D}{}\xspace}

\def\Dz      {\ensuremath{D^0}\xspace}
\def\Dzb     {\ensuremath{\Dbar^0}\xspace}
\def\DzDzb   {\ensuremath{\Dz {\kern -0.16em \Dzb}}\xspace}
\def\Dp      {\ensuremath{D^+}\xspace}
\def\Dm      {\ensuremath{D^-}\xspace}

\def\DpDm    {\ensuremath{\Dp {\kern -0.16em \Dm}}\xspace}

\def\B       {\ensuremath{B}\xspace}
\def\Bbar    {\kern 0.18em\overline{\kern -0.18em B}{}\xspace}

\def\BB      {\ensuremath{B\Bbar}\xspace} 
\def\Bz      {\ensuremath{B^0}\xspace}
\def\Bzb     {\ensuremath{\Bbar^0}\xspace}
\def\BzBzb   {\ensuremath{\Bz {\kern -0.16em \Bzb}}\xspace}
\def\Bu      {\ensuremath{B^+}\xspace}
\def\Bub     {\ensuremath{B^-}\xspace}
\def\Bp      {\ensuremath{\Bu}\xspace}
\def\Bm      {\ensuremath{\Bub}\xspace}
\def\Bpm     {\ensuremath{B^\pm}\xspace}

\def\BpBm    {\ensuremath{\Bu {\kern -0.16em \Bub}}\xspace}

\def\BorBbar    {\kern 0.18em\optbar{\kern -0.18em B}{}\xspace}
\def\DorDbar    {\kern 0.18em\optbar{\kern -0.18em D}{}\xspace}
\def\KorKbar    {\kern 0.18em\optbar{\kern -0.18em K}{}\xspace}

\mathchardef\Upsilon="7107
\def\Y#1S{\ensuremath{\Upsilon{(#1S)}}\xspace}% no space before {...}!

\mathchardef\Deltares="7101
\mathchardef\Xi="7104
\mathchardef\Lambda="7103
\mathchardef\Sigma="7106
\mathchardef\Omega="710A

\def\Deltabar{\kern 0.25em\overline{\kern -0.25em \Deltares}{}\xspace}
\def\Lbar{\kern 0.2em\overline{\kern -0.2em\Lambda\kern 0.05em}\kern-0.05em{}\xspace}
\def\Sigbar{\kern 0.2em\overline{\kern -0.2em \Sigma}{}\xspace}
\def\Xibar{\kern 0.2em\overline{\kern -0.2em \Xi}{}\xspace}
\def\Obar{\kern 0.2em\overline{\kern -0.2em \Omega}{}\xspace}
\def\Nbar{\kern 0.2em\overline{\kern -0.2em N}{}\xspace}
\def\Xb{\kern 0.2em\overline{\kern -0.2em X}{}\xspace}

\def\mes        {\mbox{$m_{\rm ES}$}\xspace}

\def\DeltaE     {\mbox{$\Delta E$}\xspace}

\newcommand{\tev}{\ensuremath{\mathrm{\,Te\kern -0.1em V}}\xspace}
\newcommand{\gev}{\ensuremath{\mathrm{\,Ge\kern -0.1em V}}\xspace}
\newcommand{\mev}{\ensuremath{\mathrm{\,Me\kern -0.1em V}}\xspace}
\newcommand{\kev}{\ensuremath{\mathrm{\,ke\kern -0.1em V}}\xspace}
\newcommand{\ev}{\ensuremath{\mathrm{\,e\kern -0.1em V}}\xspace}
\newcommand{\gevc}{\ensuremath{{\mathrm{\,Ge\kern -0.1em V\!/}c}}\xspace}
\newcommand{\mevc}{\ensuremath{{\mathrm{\,Me\kern -0.1em V\!/}c}}\xspace}
\newcommand{\gevcc}{\ensuremath{{\mathrm{\,Ge\kern -0.1em V\!/}c^2}}\xspace}
\newcommand{\mevcc}{\ensuremath{{\mathrm{\,Me\kern -0.1em V\!/}c^2}}\xspace}
\def\invfb   {\ensuremath{\mbox{\,fb}^{-1}}\xspace}

\def\mus  {\ensuremath{\rm \,\mus}\xspace}

\def\mus        {\ensuremath{\,\mu{\rm s}}\xspace}    %% microsecond
\def\ra                 {\ensuremath{\rightarrow}\xspace}
\def\to                 {\ensuremath{\rightarrow}\xspace}

\def\pep2{PEP-II}

\newcommand{\dedx}{\ensuremath{\mathrm{d}\hspace{-0.1em}E/\mathrm{d}x}\xspace}

\def\gsim{{~\raise.15em\hbox{$>$}\kern-.85em
          \lower.35em\hbox{$\sim$}~}\xspace}
\def\lsim{{~\raise.15em\hbox{$<$}\kern-.85em
          \lower.35em\hbox{$\sim$}~}\xspace}

\def\CP                {\ensuremath{C\!P}\xspace}
\xspace

\newcommand{\jprlBase}       {Phys.\ Rev.\ Lett.\xspace}
\newcommand{\jprBase}        {Phys.\ Rev.\xspace}
\newcommand{\jplBase}        {Phys.\ Lett.\xspace}
\newcommand{\nimBaseA}       {Nucl.\ Instr.\ Methods Phys.\ Res., Sect.\ A\xspace}

\newcommand{\nima}      [1]  {\nimBaseA~{\bf #1}}

\newcommand{\plb}       [1]  {\jplBase\ B~{\bf #1}}

\newcommand{\jprl}      [1]  {\jprlBase\ {\bf #1}}
\newcommand{\jprd}      [1]  {\jprBase\ D~{\bf #1}}
\def\jetset74   {\mbox{\tt Jetset \hspace{-0.5em}7.\hspace{-0.2em}4}\xspace}

\def\Acp    {\ensuremath{{\cal A}_{\CP }}\xspace}
\def\Acpp   {\ensuremath{{\cal A}_{\CP+ }}\xspace}
\def\Acpm   {\ensuremath{{\cal A}_{\CP- }}\xspace}
\def\Acppm  {\ensuremath{{\cal A}_{\CP \pm }}\xspace}
\def\Rcp    {\ensuremath{{\cal R}_{\CP }}\xspace}
\def\Rcpp   {\ensuremath{{\cal R}_{\CP+ }}\xspace}
\def\Rcpm   {\ensuremath{{\cal R}_{\CP- }}\xspace}
\def\Rcppm  {\ensuremath{{\cal R}_{\CP \pm }}\xspace}
\def\Aads   {\ensuremath{{\cal A}_{ADS }}\xspace}
\def\Rads   {\ensuremath{{\cal R}_{ADS }}\xspace}
\def\cpp    {\ensuremath{\CP+ }\xspace}
\def\cpm    {\ensuremath{\CP- }\xspace}

\def\de {\ensuremath{{\rm \Delta}E}\xspace}

\def\figurebox#1#2#3{%
    \def\arg{#3}%
    \ifx\arg\empty
    {\hfill\vbox{\hsize#2\hrule\hbox to #2{\vrule\hfill\vbox to #1{\hsize#2\vfill}\vrule}\hrule}\hfill}%
    \else
    {\hfill\epsfbox{#3}\hfill}%
    \fi}

\def\acppc{0.09} 
\def\acppstat{0.13} 
\def\acppsys{0.06}
\def\acpmc{-0.23} 
\def\acpmstat{0.21}
\def\acpmsys{0.07}

\def\aadsc{-0.34}
\def\aadsstat{0.43}
\def\aadssys{0.16}

\def\rcppc{2.17}
\def\rcppstat{0.35}
\def\rcppsys{0.09} 
\def\rcpmc{1.03} 
\def\rcpmstat{0.27}
\def\rcpmsys{0.13}

\def\radsc{0.066} 
\def\radsstat{0.031}
\def\radssys{0.010} 

\newcommand{\BaBarYear}      {09}
\newcommand{\BaBarNumber}    {019}
\newcommand{\SLACPubNumber}  {13793}

\begin{document}
\noindent
 \babar-PUB-\BaBarYear/\BaBarNumber \\
 SLAC-PUB-\SLACPubNumber \\

\title{Measurement of $CP$ violation observables and parameters for the decays \boldmath{$B^{\pm}\to DK^{*\pm}$}}

%
%% author list as of 05-May-2009 (488 authors)
%
\author{B.~Aubert}
\author{Y.~Karyotakis}
\author{J.~P.~Lees}
\author{V.~Poireau}
\author{E.~Prencipe}
\author{X.~Prudent}
\author{V.~Tisserand}
\affiliation{Laboratoire d'Annecy-le-Vieux de Physique des Particules (LAPP), Universit\'e de Savoie, CNRS/IN2P3,  F-74941 Annecy-Le-Vieux, France}
\author{J.~Garra~Tico}
\author{E.~Grauges}
\affiliation{Universitat de Barcelona, Facultat de Fisica, Departament ECM, E-08028 Barcelona, Spain }
\author{M.~Martinelli$^{ab}$}
\author{A.~Palano$^{ab}$ }
\author{M.~Pappagallo$^{ab}$ }
\affiliation{INFN Sezione di Bari$^{a}$; Dipartimento di Fisica, Universit\`a di Bari$^{b}$, I-70126 Bari, Italy }
% ...
\collaboration{The \babar\ Collaboration}
\noaffiliation

\begin{abstract}
We study the decay $B^-\to DK^{*-}$ using a sample of 379$\times 10^6$ $\Upsilon(4S)\to$ \BB events collected 
with the \babar\  detector at the PEP-II $B$-factory. We perform a ``GLW'' analysis where the $D$ meson decays 
into either a $CP$-even ($CP+$) eigenstate ($K^+K^-$, $\pi^+\pi^-$), $CP$-odd ($CP-$) eigenstate ($K^0_S\pi^0$, $K^0_S\phi$, 
$K^0_S\omega$) or a non-$CP$ state ($K^-\pi^+$). We also analyze $D$ meson decays into $K^+\pi^-$ from a Cabibbo-favored 
$\overline{D}^0$ decay or doubly suppressed $D^0$ decay (``ADS'' analysis). We measure observables 
that are sensitive to the CKM angle $\gamma$: the partial-rate charge asymmetries \Acppm, 
the ratios  \Rcppm of the $B$-decay branching fractions in $CP\pm$ and non-$CP$ decay, 
the ratio \Rads of the charge-averaged branching fractions, and the charge asymmetry 
\Aads of the ADS decays:
  $\Acpp= \acppc \pm \acppstat \pm \acppsys$, 
  $\Acpm= \acpmc \pm \acpmstat \pm \acpmsys$, 
  $\Rcpp =\rcppc \pm \rcppstat \pm \rcppsys$,
  $\Rcpm = \rcpmc \pm \rcpmstat \pm \rcpmsys$,
  $\Rads =\radsc \pm \radsstat \pm \radssys$, and
  $\Aads = \aadsc \pm \aadsstat \pm \aadssys$,
where the first uncertainty is statistical and the second is systematic. 
Combining all the measurements and using a frequentist approach
yields the magnitude of the ratio between the Cabibbo-suppressed and favored amplitudes, 
$r_B$ = 0.31 with a one (two) sigma confidence
level interval
of [0.24, 0.38] ([0.17, 0.43]). The value $r_B=0$ is excluded at the 3.3 sigma level. 
A similar analysis
excludes values of $\gamma$ in the intervals
$[0,~7]^{\circ},~[55,~111]^{\circ}$, and $[175,~180]^{\circ}$ ([85, 99]$^{\circ}$) 
at the one (two) sigma confidence level.
\end{abstract}

% PACS, the Physics and Astronomy Classification Scheme.

\maketitle

%*******************************
\section{Introduction}
%*******************************
The Standard Model accommodates $CP$ violation through a single phase in the Cabibbo-Kobayashi-Maskawa (CKM) quark 
mixing matrix~$V$~\cite{CKM}. The self consistency of this mechanism can be tested by over-constraining the associated 
unitarity triangle~\cite{Jarl,pdg} using many different measurements, mostly involving decays of $B$ mesons. In this paper we 
concentrate on the angle $\gamma\,\equiv$  arg$(-V_{ud}V^*_{ub}/V_{cd}V^*_{cb})$ by studying $B$ meson decay channels where 
$b\to c\overline{u}s$ and $b\to u\overline{c}s$ tree amplitudes interfere. We use two techniques, one suggested by Gronau and 
London~\cite{GL} and Gronau and Wyler~\cite{GW} (GLW) and the other suggested by Atwood, Dunietz and Soni~\cite{ADS} (ADS) 
to study $\gamma$. Both techniques rely on final states that can be reached from both $D^0$ and \Dzb decays. 
As discussed in Ref.~\cite{AbiGLWADS} the combination of the GLW and ADS observables can be very useful in resolving certain ambiguities
inherent in each of the techniques. In this paper we use the decay  $B^-\to DK^{*-}(892)$~\cite{892}  to measure the GLW and
ADS observables.

In the GLW analysis the $D$ meson~\cite{D0D0bar} from $B^-\to DK^{*-}$ decays into either a $CP$-even ($CP+$) 
eigenstate ($K^+K^-$, $\pi^+\pi^-$) or a $CP$-odd ($CP-$) eigenstate ($K_s\pi^0,~K_s\phi,~K_s\omega$). The size of the 
interference between the two competing amplitudes depends on the CKM angle $\gamma$ as well as other parameters that are 
$CP$-conserving, discussed below. References~\cite{GL,GW} define several observables that depend on measurable quantities:
\begin{small}
\begin{eqnarray}
  \Rcppm &=&2\frac{\Gamma(B^-\to D^0_{CP^{\pm}}K^{*-})+\Gamma(B^+\to  D^0_{CP^{\pm}}K^{*+})}
  {\Gamma(B^-\to D^0_{K\pi} K^{*-})+\Gamma(B^+\to \Dzb_{K\pi}K^{*+})}, \nonumber \\
  \Acppm &=&\frac{\Gamma(B^-\to D^0_{CP^{\pm}}K^{*-})-\Gamma(B^+\to D^0_{CP^{\pm}}K^{*+})}
  {\Gamma(B^-\to D^0_{CP^{\pm}}K^{*-})+\Gamma(B^+\to D^0_{CP^{\pm}}K^{*+})}. \nonumber
\end{eqnarray}
\end{small}

\noindent
Here $D^0_{CP\pm}$ refers to a neutral $D$ meson decaying into either a $CP+$ or $CP-$ eigenstate. 
%% put this into the reference...We require $K^*$ to decay only into $\KS\pi$.

\Rcppm and \Acppm depend on the physical parameters as follows:
\begin{eqnarray}
  \Rcppm &=&1+r_B^2\pm 2r_B\cos\delta_B\cos\gamma, \label{eq:R} \\
  \Acppm &=&\pm 2r_B\sin\delta_B\sin\gamma/\Rcppm. \label{eq:A}
\end{eqnarray}

\noindent
Here $r_B$ is the magnitude of the ratio of the suppressed $B^-\to \Dzb K^{*-}$ 
and favored $B^-\to D^0 K^{*-}$ decay amplitudes, respectively, and  $\delta_B$ is the $CP$-conserving
% strong 
phase difference 
between these amplitudes.
In this analysis we neglect the effects of $CP$ violation in $D$ meson
 decays, and, as justified in Ref.~\cite{Abi}, the very small effect of \DzDzb mixing.

\Rcppm is calculated using:
\begin{equation} \label{equation:Rcal}
{\cal R}_{CP\pm}=\frac{N_{CP\pm}}{N_{non-CP}}\times \frac{\epsilon_{non-CP}}{\epsilon_{CP\pm}}
\end{equation}

\noindent
where $N_{CP\pm}$ and $N_{non-CP}$ are the event yields for the $CP$ 
and non-$CP$ modes, respectively and  $\epsilon_{non-CP}$ and $\epsilon_{CP\pm}$
are correction
factors that depend on branching fractions and reconstruction efficiencies.
\Acppm is calculated using the event yields split by
the charge of the $B$ meson.

We define two additional quantities whose experimental estimators are normally distributed even
when the value of
$r_B$ is     comparable to its uncertainty:
\begin{eqnarray}\label{xvariables}
  x_{\pm}&=& r_B\cos(\delta_B\pm\gamma)\\
         &=& \frac{R_{CP+}(1 \mp \Acpp) - R_{CP-}(1 \mp \Acpm)}{4}. \nonumber
\end{eqnarray}
Since $x_{\pm}$ are also directly measured in Dalitz-plot analyses \cite{Giri:2003ty}, the different results can be 
compared and combined with each other. We note that an additional set of quantities measured in Dalitz-plot analyses, 
$y_{\pm} = r_B\sin(\delta_B\pm \gamma)$, are not directly accessible through the GLW analysis.

In the ADS technique, $B^-\to DK^{*-}$ decays to $[K^+\pi^-]_DK^{*-}$, where  $[K^+\pi^-]_D$ 
indicates that these particles are neutral $D$ meson ($\Dz$ or $\Dzb$) decay products. This final state can be reached 
from $B\to D^0K^{*-}$ and the doubly-Cabibbo-suppressed decay $D^0\to K^+\pi^-$ or $B^-\to \Dzb K^{*-}$ followed by the 
Cabibbo-favored decay $\Dzb\to K^+\pi^-$. In addition, the final state $[K^-\pi^+]_DK^{*-}$ is used
for normalization. We label the decays where the $K$ and $K^*$ have the same (opposite) charge as ``right (wrong) sign''
where the labels reflect that one mode occurs much more often than the other.  

In analogy with the GLW method we define two measurable quantities, \Rads and \Aads, as follows:
\begin{footnotesize}
\begin{eqnarray}
  \Rads &=&\frac{\Gamma(B^-\to[K^+\pi^-]_DK^{*-})+\Gamma(B^+\to[K^-\pi^+]_DK^{*+})}
  {\Gamma(B^-\to [K^-\pi^+]_DK^{*-})+\Gamma(B^+\to [K^+\pi^-]_DK^{*+})}, \nonumber \\
  \Aads &=&\frac{\Gamma(B^-\to[K^+\pi^-]_DK^{*-})-\Gamma(B^+\to[K^-\pi^+]_DK^{*+})}
  {\Gamma(B^-\to[K^+\pi^-]_DK^{*-})+\Gamma(B^+\to[K^-\pi^+]_DK^{*+})}. \nonumber
\end{eqnarray}
\end{footnotesize}
\noindent
\Rads and \Aads are related to physically interesting quantities by:
\begin{eqnarray}
  \Rads &=&r_D^2+r_B^2+2r_Dr_B\cos(\delta_B+\delta_D) \cos\gamma, \label{eq:Rads} \\
  \Aads &=&2r_Dr_B\sin(\delta_B+\delta_D)\sin\gamma/\Rads. \label{eq:Aads}
\end{eqnarray}
Here $r_D$ is the magnitude of the ratio of suppressed $D^0\to K^+\pi^-$ and favored $D^0\to K^-\pi^+$ 
decay amplitudes, respectively, while $\delta_D$ is the $CP$-conserving strong phase difference between these two amplitudes.
Both  $r_D$ and $\delta_D$ have been measured and we use the values given in Ref.~\cite{HFAG2008}: $r_D=0.0578\pm 0.0008$ and
$\delta_D=21.9^{+11.3}_{-12.4}$ degrees. Estimates for $r_B$ are in the range 
$0.1\le r_B\le 0.3$~\cite{rb,BAD1141}.

It has been pointed out in Ref.~\cite{rb} that complications due to possible variations in $r_B$ and/or $\delta_B$
as a result of the finite width of a resonance such as the $K^*$ and its overlap with other states 
can be taken into account using an alternate formalism. However, in this paper we
choose to follow the procedures in Refs.~\cite{BAD1141, BAD697} and incorporate the effects of the non-$K^*$ $DK\pi$ events
and finite width of the $K^*$ into the systematic uncertainties of our ${\cal A}$ and ${\cal R}$ measurements.

%*******************************
\section{\boldmath The \babar\ detector and dataset}
%*******************************
The \babar\ detector has been described in detail in Ref.~\cite{babardet} and therefore will only be briefly 
discussed here. The trajectories of charged tracks are measured with a  five-layer double-sided silicon vertex 
tracker (SVT) and a 40-layer drift chamber (DCH). Both the SVT and DCH are located inside 
a 1.5 T magnetic field. Photons are detected by means of a CsI(Tl) crystal calorimeter also 
located inside the magnet. Charged particle identification is determined from information provided by a 
ring-imaging Cherenkov device (DIRC) in combination with ionization measurements (\dedx) from the 
tracking detectors. The \babar\ detector's response to various physics processes as well as varying beam 
and environmental conditions is modeled with simulation software based on the {\sc Geant}4~\cite{geant4} toolkit.
We use {\sc EvtGen}~\cite{evtgen} to model the kinematics of $B$ mesons and {\sc Jetset}~\cite{jetset}  
to model continuum processes ($e^+e^-\to c\overline{c},~u\overline{u},~d\overline{d},~s\overline{s}$).
This analysis uses data collected at and near the $\Upsilon(4S)$ resonance with the \babar\ detector
at the PEP-II storage ring. The data set consists of 345 $\invfb$ collected at the peak of the $\Upsilon(4S)$
(379 $\times 10^6$ \BB pairs) and 35~$\invfb$ collected 40~\mev below the resonance peak~(off-peak data).

This analysis is a combined update of the previous $\babar$  ADS~\cite{BAD1141} and GLW~\cite{BAD697} studies 
of $B^-\to DK^{*-}$, which used 232 $\times 10^6$~\BB pairs. Other new features 
in this analysis include the improvement in background suppression, the refinement of various candidate selection criteria, 
and an update of the estimation of systematic uncertainties. The major change is the choice of neural networks in the GLW analysis 
over Fisher discriminants, which were used in the previous analysis. We verify the improvements on both signal 
efficiency and continuum background 
rejection in the GLW decay channels with simulated signal and continuum  events. The increases in signal 
efficiency range from 3\% to 14\% for all channels except $\KS\phi$, which has the same efficiency. For continuum 
suppression, the neural networks perform 10\% to 57\% better across all channels except $K^+K^-$, which displays the 
same performance.

%*******************************
\section{\boldmath The GLW Analysis}\label{GLW}
%*******************************
We reconstruct $\Bm\to D\Kstarm$  candidates with the subsequent decays $\Kstarm\to\KS\pim$, 
$\KS\to\pip\pim$  and with the $D$ meson decaying into six decay final states: $\Dz\to\Km\pip$ (non-\CP final state); \KpKm, 
$\pip\pim$ (\cpp eigenstates); and $\KS\piz$, $\KS\phi$, $\KS\omega$ (\cpm eigenstates). We optimize our event 
selection criteria by maximizing the figure of merit $S/\sqrt{S+B}$, with $S$ the number of signal events and $B$ the 
number of background events, determined for each channel using simulated signal and background 
events. Kaon and pion candidates (except for the pions from \KS decays)  are selected 
using a likelihood-based particle identification algorithm which 
relies on \dedx information measured in the DCH and the SVT, and Cherenkov photons in the DIRC. The efficiencies of 
the selectors are typically above 85\% for momenta below 4 $\gev$ while the kaon and pion misidentification
 rates are at the few 
percent level for particles in this momentum range.

The \KS candidates are formed from oppositely charged tracks assumed to be pions with a reconstructed invariant 
mass within 13~\mevcc\ (four standard deviations) of the known \KS mass~\cite{pdg}, $m_{\KS}$. 
All \KS candidates are refitted so that their invariant mass equals $m_{\KS}$ (mass constraint). They are also 
constrained to emerge from a single vertex (vertex constraint). For those retained to build a \Kstarm candidate, we further 
require that their flight direction and length be consistent with a \KS\ coming from the interaction point. 
The \KS\ candidate flight path and momentum vectors must make an acute angle and the flight length in the 
plane transverse to the beam direction must exceed its uncertainty by three standard deviations. 
\Kstarm candidates are formed from a \KS and a charged particle with a vertex constraint. 
We select \Kstarm candidates that have an invariant mass within 75~\mevcc of the known mean value for a 
$K^*$~\cite{pdg}. Finally, since the \Kstarm in $\Bm\to D\Kstarm$ is longitudinally polarized, we require 
$|\cos\theta_{H}| \geq 0.35$, where $\theta_{H}$ is the angle in the \Kstarm\ rest frame between the daughter 
pion momentum and the parent \B momentum. The helicity distribution discriminates well between a \B-meson decay 
and a false \B-meson candidate from the 
%\epem \to \qqbar\ ($q\in \{u,d,s,c\}$) 
continuum, since the former is distributed as $\cos^2\theta_{H}$ and the latter has an approximately flat distribution.

Some decay modes of the $D$ meson contain a neutral pion. We combine pairs of photons to form  \piz candidates with a total 
energy greater than 200~\mev and an invariant mass between 115 and 150~\mevcc. A mass constrained fit is applied 
to the selected \piz\ candidate momenta. Composite particles ($\phi$ and $\omega$) included in the \cpm\  modes are 
vertex-constrained. Candidate $\phi$ ($\omega$) mesons are constructed from $\Kp\Km$ ($\pip\pim\piz$) particle 
combinations with an invariant mass required to be within two standard deviations, corresponding to   12 (20)~\mevcc, from the 
known peak values~\cite{pdg}. Two further requirements are made on the $\omega$ candidates. The magnitude of the 
cosine of the helicity angle $\theta_H$ between the $D$ momentum in the rest frame of the $\omega$ and the normal to the 
plane containing all three decay pions must be greater than 0.35 (this variable has a $\cos^2\theta_H$ distribution for 
signal candidates and is approximately flat for background). The Dalitz angle~\cite{dalitzomega} $\theta_D$ is defined as the 
angle between the momentum of one daughter pion in the $\omega$ rest frame and the direction of one of the other 
two pions in the rest frame of the two pions. For signal candidates, the cosine of the Dalitz angle follows a 
$\sin^2\theta_D$ distribution,
 while it is approximately flat for the background. Therefore we require the cosine of the Dalitz angle 
of signal candidates to have a magnitude smaller than~0.8.

All $D$ candidates are mass constrained and, with the
exception of the $\KS\piz$ final state, vertex constrained. We select $D$ candidates 
with an invariant mass differing from the known mass~\cite{pdg} by less than 12~\mevcc for all channels except 
$\KS\piz$ (30~\mevcc) and $\KS\omega$ (20~\mevcc). These limits are about twice the corresponding RMS mass resolutions.

Suppression of backgrounds from %$e^+e^-\to q\overline{q}$ 
continuum events is achieved by using event-shape and angular variables. 
The $B$ meson candidate is required to have $|\cos\theta_T| \le 0.9$, where $\theta_T$ is 
the angle between the thrust axis of the $B$ meson and that of the rest of the event. The distribution of 
$|\cos\theta_T|$ is uniform in $B\Bbar$ events and strongly peaked near 1 for continuum events.

\begin{figure}[ht]
  \begin{center}
    \includegraphics[width=0.45\linewidth]{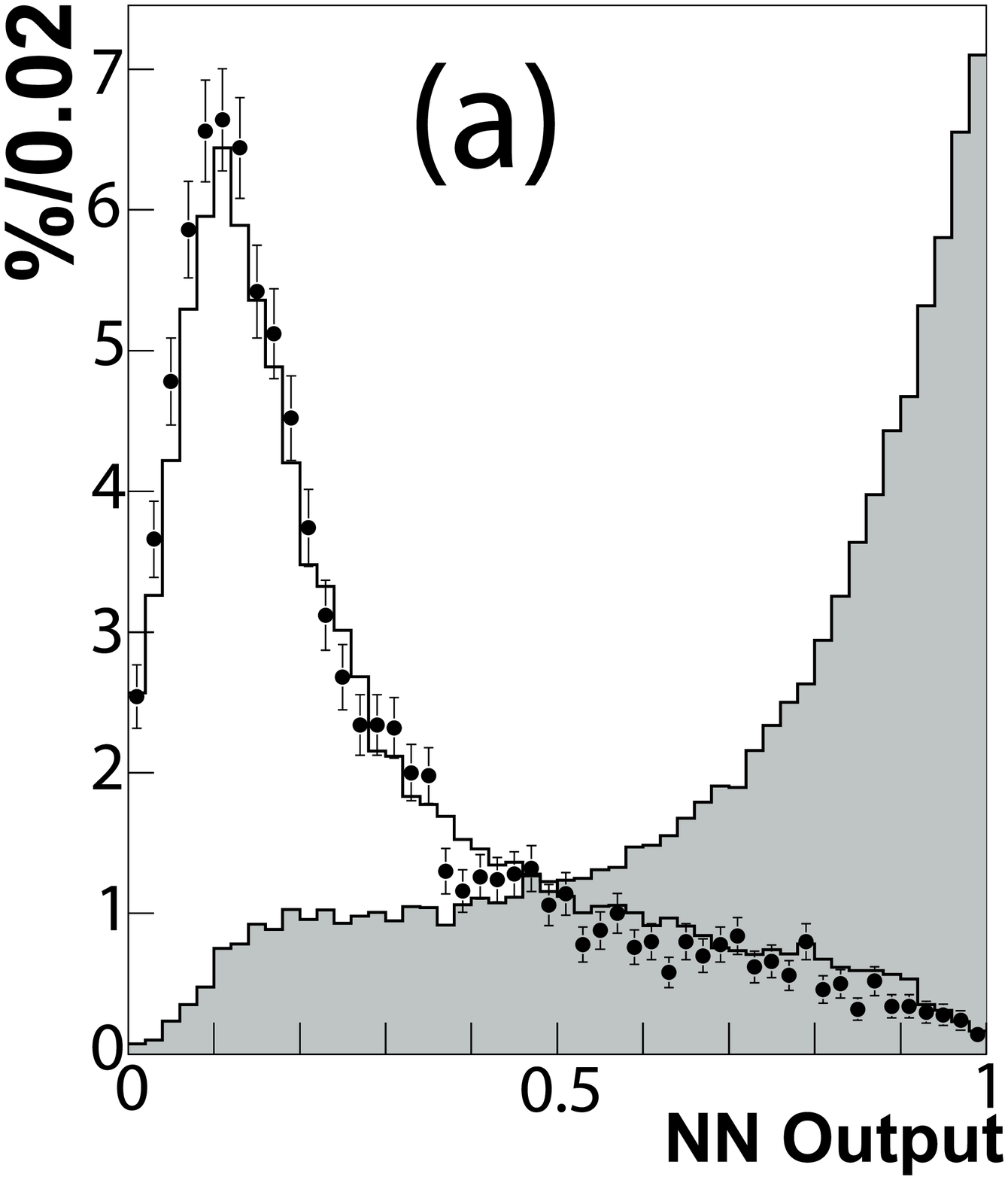}
    \includegraphics[width=0.45\linewidth]{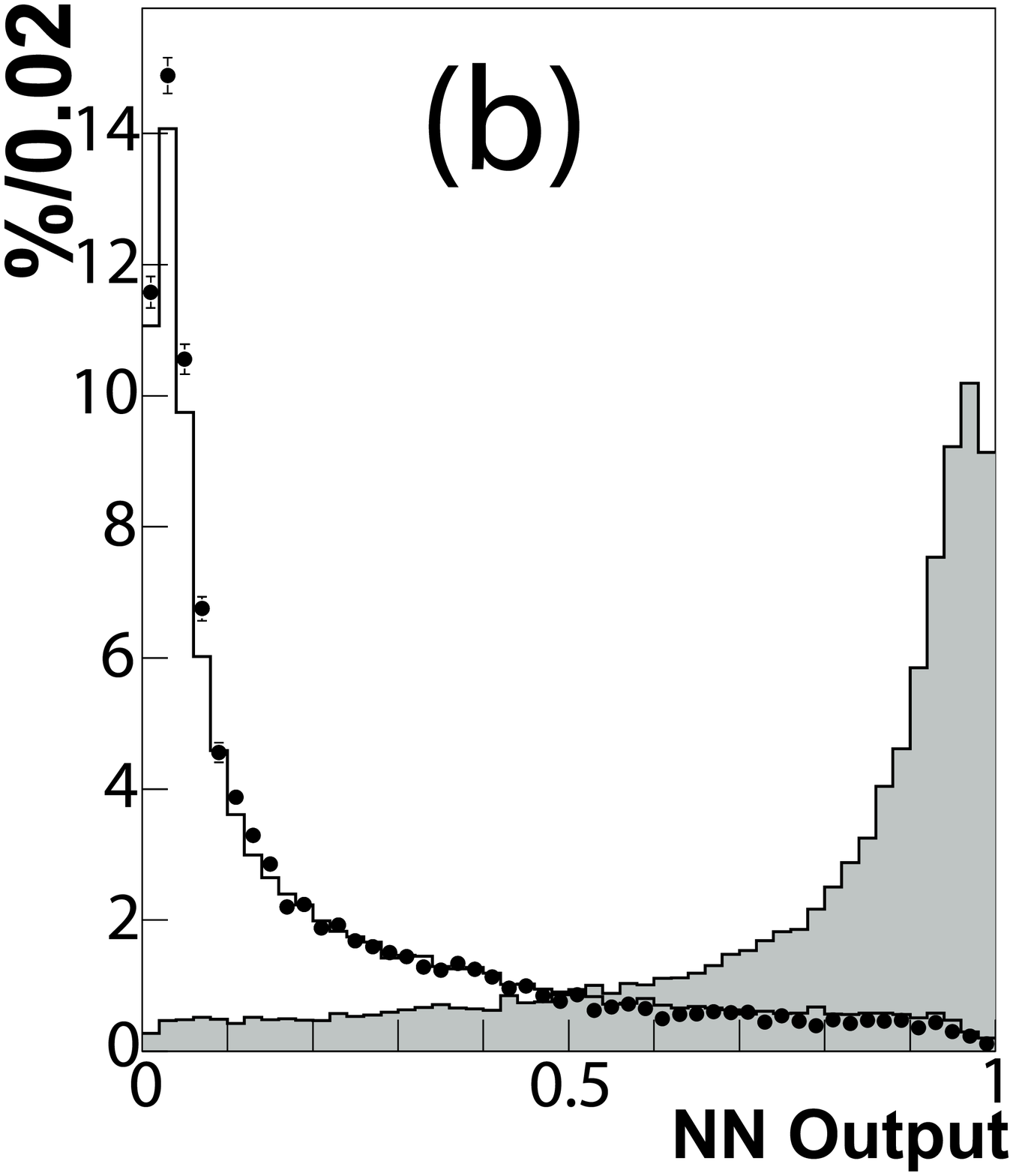}
    \includegraphics[width=0.45\linewidth]{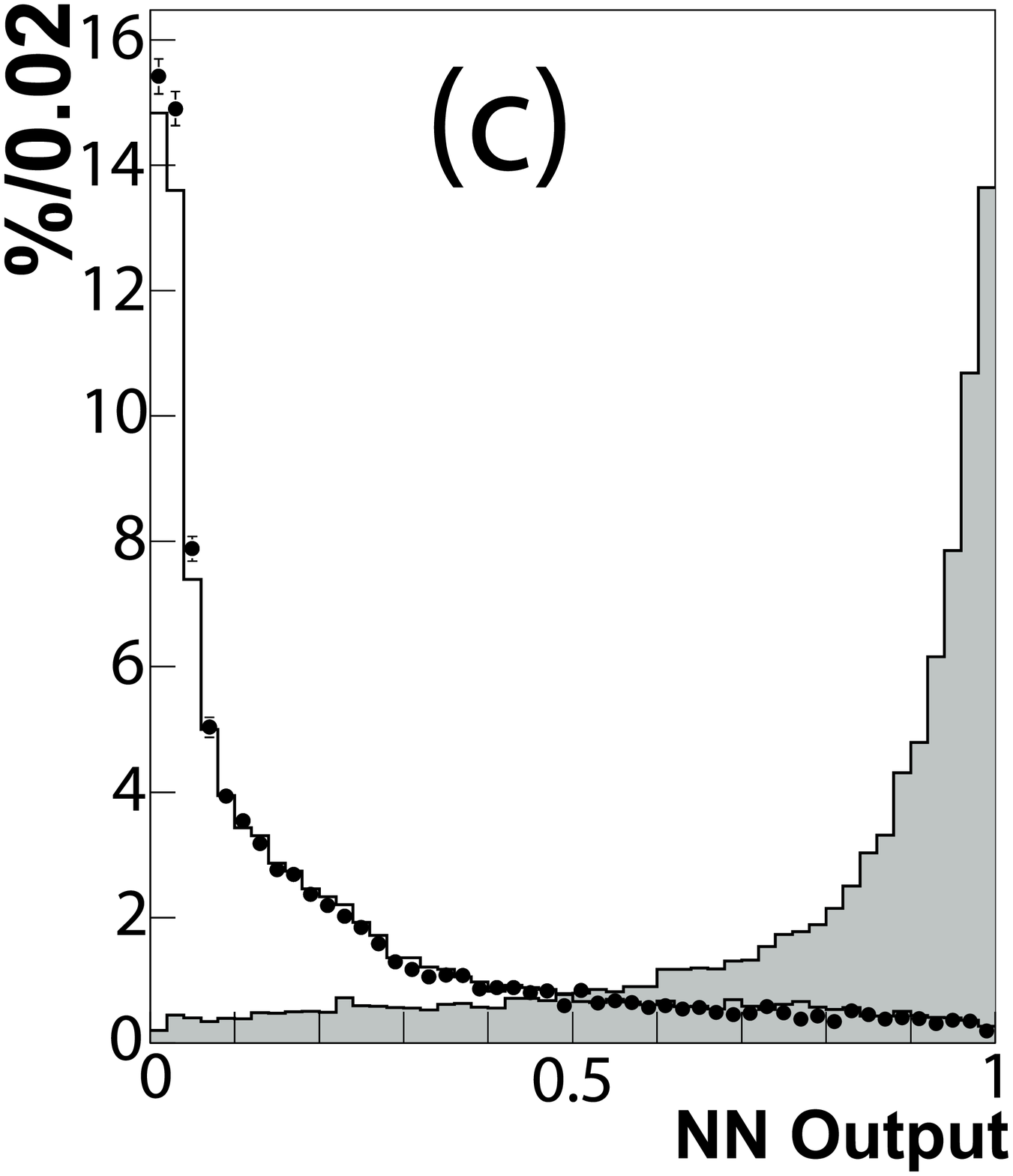}
 \includegraphics[width=0.45\linewidth]{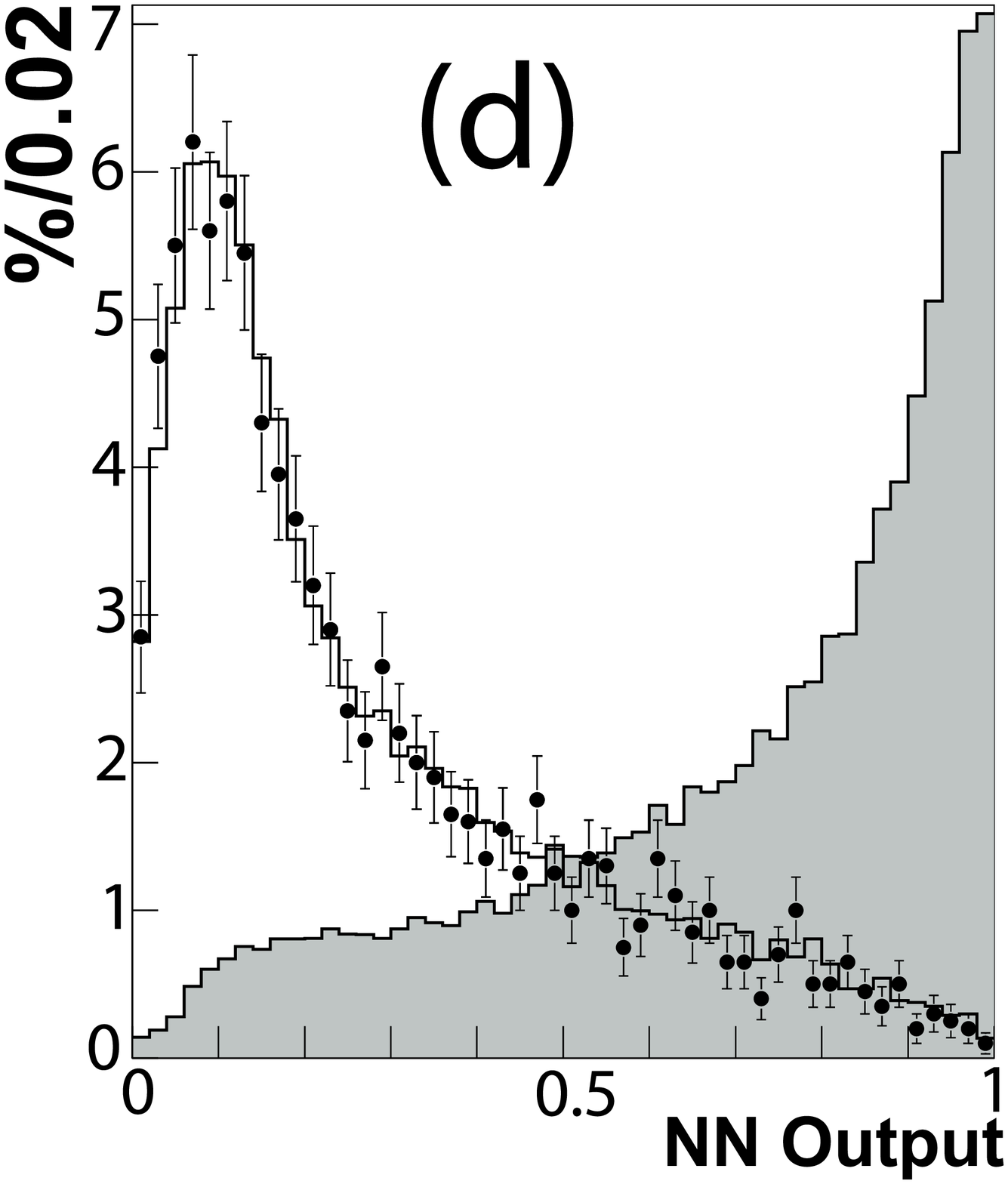}
    \includegraphics[width=0.45\linewidth]{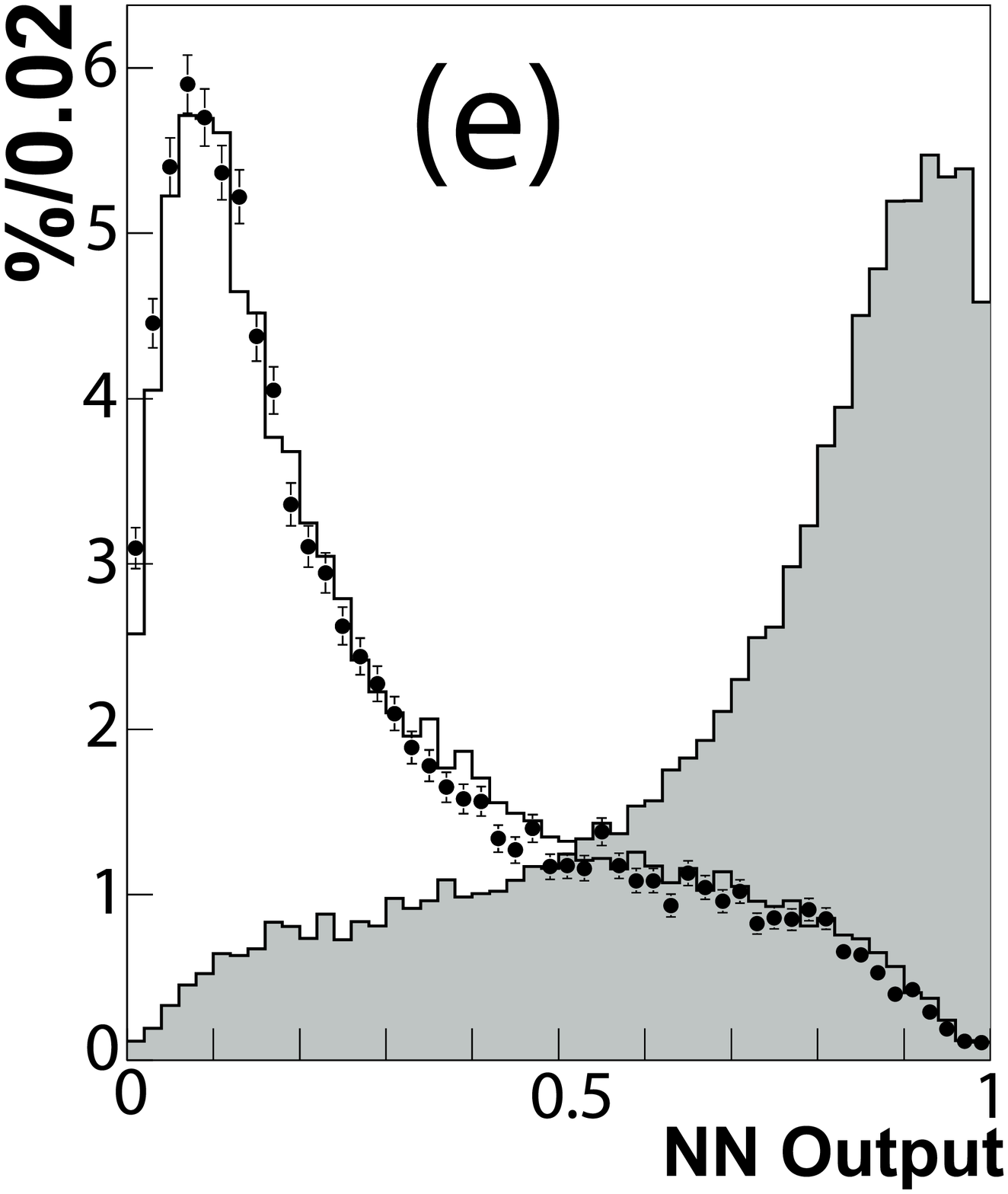}
    \includegraphics[width=0.45\linewidth]{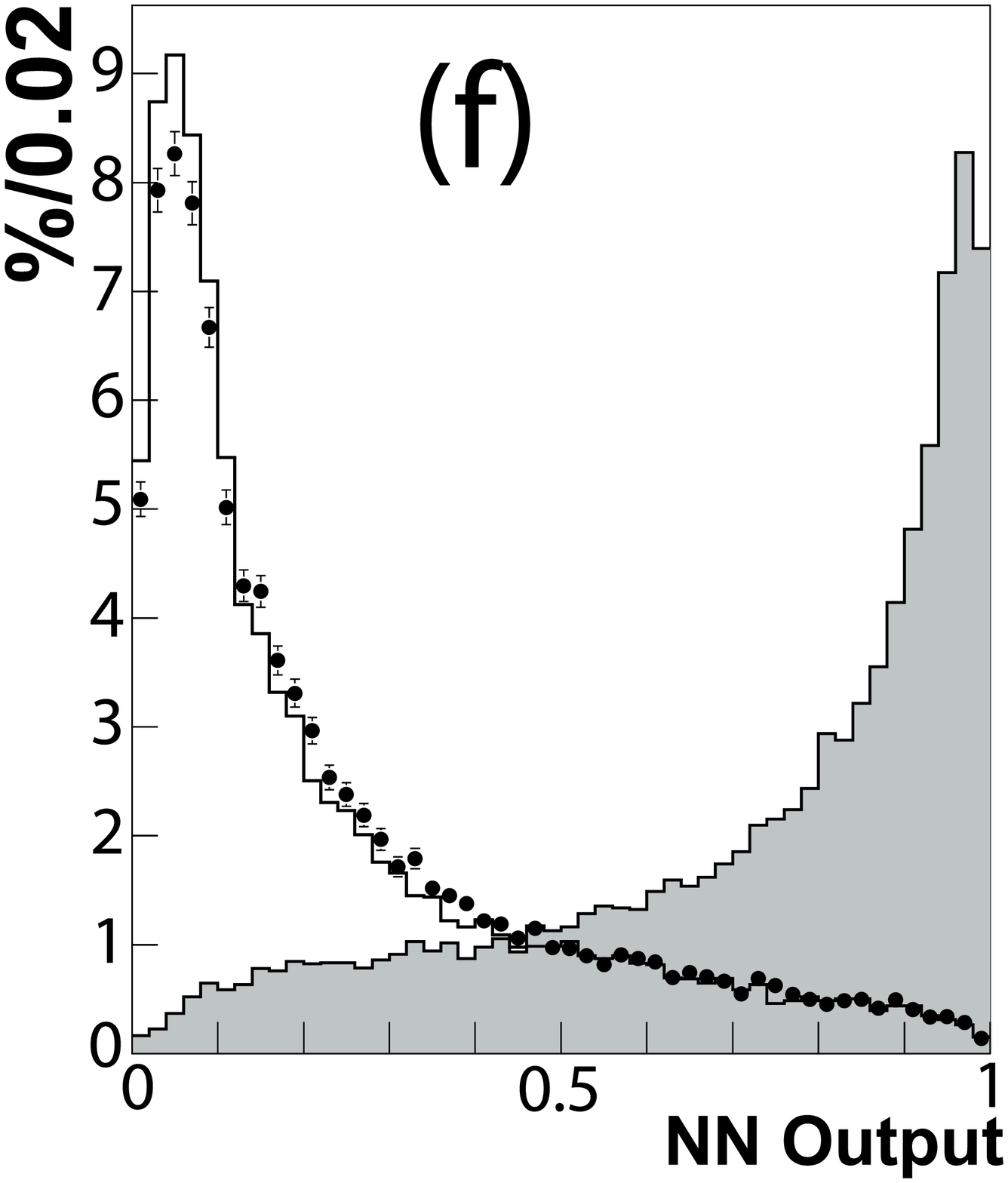}
    \caption{\label{fig:GLWnnPerformance}
      Neural network (NN) outputs and results of the NN verifications of (a) $K^+K^-$, (b) $\pip\pim$, 
      (c) $\KS\piz$, (d) $\KS\phi$, (e) $\KS\omega$, and (f) $K^-\pip$ subsamples of the GLW analysis. 
      The samples used to produce the output are shown as histograms. 
      The signal (Monte Carlo simulation) is the shaded histogram peaking near 1;
 the continuum (Monte Carlo simulation) is the histogram peaking near 0.
      The off-peak data 
used to check the NN are overlaid as data points.}
  \end{center}
\end{figure}

A neural network (NN) is used to further reduce the $e^+e^-\to\qqbar$ ($q = u, d, s, c$) contribution to our data sample.
Seven variables are used in the NN with three being the angular moments $L_0$, $L_1$ and $L_2$. These 
moments are defined by $L_j=\sum_{i}p^*_i|\cos\theta^*_i|^j$ where the sum is over charged and neutral particles
not associated with the $B$-meson candidate. Here $p^*_i$ ($\theta^*_i$) is the momentum (angle) of the $i$th 
particle with respect to the thrust of the candidate $B$ meson in the center-of-mass (CM) frame. Additional 
details on the moments can be found in Ref.~\cite{muriel}. The NN also uses the ratio $R_2=H_2/H_0$ of Fox-Wolfram 
moments~\cite{foxwolfram}, the cosine of the angle between the $B$ candidate momentum vector and the beam axis 
($\cos\theta_B$), $\cos\theta_T$ (defined above), and the cosine of the angle between a $D$ daughter momentum vector 
in the $D$ rest frame and the direction of the $D$ in the \B meson rest frame ($\cos\theta_H(D)$). The 
distributions of all the above variables show distinct differences between signal and continuum events and thus can 
be exploited by a NN to preferentially select $B\Bbar$ events. Each decay mode has its own unique NN trained with signal and continuum 
Monte Carlo events. After training, the NNs are 
then fed with independent sets of signal and continuum Monte Carlo events to produce 
NN outputs for each decay mode. Finally, we verify that the NNs have consistent outputs for off-peak data 
(continuum data collected below the $\Upsilon(4S)$) and $q\overline{q}$ Monte Carlo events. The 
separations between signal and continuum background are shown in Fig.~\ref{fig:GLWnnPerformance}. We 
select candidates with neural network output above 0.65 ($K^+K^-$), 0.82 ($\pip\pim$), 0.91 ($\KS\piz$), 
0.56 ($\KS\phi$), 0.80 ($\KS\omega$), and 0.73 ($K^-\pip$). 
Our event selection is optimized to maximize the  
significance of the signal yield, determined using simulated signal and background events.

We identify \B candidates using two nearly independent kinematic variables: the beam-energy-substituted mass
$\mes=\sqrt{(s/2+{{\bf p}_0 \cdot {\bf p}_B})^2/E_0^2-p_B^2}$ and the energy difference $\Delta E=E_B^*-\sqrt{s}/2$, where 
$E$ and $p$ are energy and momentum. The subscripts 0 and $B$ refer to the \epem-beam system and the \B candidate, 
respectively; $s$ is the square of the CM energy and the asterisk labels the CM frame. The \mes\ distributions are 
all described by a Gaussian function $\mathcal{G}$ centered at the \B mass with a resolution (sigma) of 2.50, 2.55, and 2.51~\mevcc 
for the \cpp, \cpm and non-$CP$ mode, respectively. The \de distributions are centered on zero for signal 
with a resolution of 11 to 13~\mev\ for all channels except $\KS\piz$ for which the resolution is asymmetric and is about 
30~\mev. We define a signal region through the requirement $|\Delta E| <$ 50 (25)~\mev\ for $\KS\piz$ (all other modes).

A potentially dangerous background for the $\Bm\to D(\pip\pim)\Kstarm(\KS\pim)$ channel is the decay mode $\Bm\to D(\KS\pip\pim)\pim$ 
which contains the same final-state particles as the signal but has a branching fraction 600 times larger. We therefore explicitly 
veto any selected \B~candidate containing a $\KS\pip\pim$ combination within 60~\mevcc~of the \Dz~mass.
%The remaining \BB background is negligible.

The fraction of events with more than one acceptable \B candidate depends on the $D$ decay mode and is always less
than 8\%.
To select the best \B candidate in those events where we find more than one acceptable candidate,
 we choose the one with the smallest $\chi^2$ formed from the differences of the measured and world average \Dz 
and $\Kstarm$ masses divided by the mass spread that includes the resolution and, for the \Kstarm, the natural width: 
\begin{eqnarray}\label{equation:multiChi2}
  \chi^2 &=& \chi^2_{M_{D^0}} + \chi^2_{M_{K^{*-}}}\\ \nonumber
         &=& \frac{(M_{D^0} - M^{PDG}_{D^0})^2}{\sigma^2_{M_{D^0}}} + 
           \frac{(M_{K^{*-}} - M^{PDG}_{K^{*-}})^2}{\sigma^2_{M_{K^{*-}}}+{\Gamma^2_{K^{*-}}/c}^4}.
\end{eqnarray}
Simulations show that negligible bias is introduced by this choice and the correct candidate is picked at least 86\% of the time.

From the simulation of signal events, the total reconstruction efficiencies are: 12.8\% and 12.3\% for the \cpp 
modes $D\to\KpKm$  and $\pip\pim$; 5.6\%, 8.9\%, and 2.4\%  for the \cpm modes $D\to\KS\piz$, $\KS\phi$ and 
$\KS\omega$; 12.8\% for the non-\CP mode $\Dz\to\Km\pip$.

To study \BB\ backgrounds we look in sideband regions in \de and $m_{D}$. 
We define the \de\ sideband in the interval $60 \leq \de \leq 200 \mev $ for all modes. This region is used to determine the 
combinatorial background shapes in the signal and $m_{D}$ sideband. We choose not to use a 
lower sideband because of the $D^*K^*$ backgrounds in that region. The sideband region in 
$m_{D}$ is slightly mode dependent with a typical requirement that $m_D$ differs from the \Dz mass by more than four 
and less than 10 standard 
deviations. This region provides sensitivity to background sources which mimic signal both in \de\ and \mes 
and originate from either charmed or charmless \B meson decays that do not contain a true $D$ meson. As many of 
the possible contributions to this background are not well known, we measure its size by including the 
$m_{D}$ sideband in the fit described below.

An unbinned extended maximum likelihood fit to the \mes distributions of selected $B$ candidates in the range 
$5.2\leq\mes\leq 5.3$~\gevcc is used to 
determine signal and background yields. We use the signal yields to calculate the \CP-violating quantities \Acp and \Rcp. 
We use the same mean and width
of the Gaussian function $\mathcal{G}$ to 
describe the signal shape for all modes considered. The combinatorial background in the \mes\ distribution is 
modeled with the so-called ``ARGUS'' empirical threshold function $\mathcal{A}$~\cite{argus}. It is defined as:
\begin{equation} \label{equation:ARGUS}
  \mathcal{A} (\mes) \propto \mes \sqrt{1-x^2}\,{\rm exp}^{-\xi (1-x^2)},
\end{equation}
where $x= \mes / E_{\rm max}$ and $E_{\rm max}$ is the maximum mass for pair-produced $B$ mesons given the collider beam energies and is 
fixed in the fit at 5.291$\gev/c^2$. The ARGUS shape is governed by one parameter $\xi$ that is 
left free in the fit. 
We fit simultaneously \mes distributions of nine samples: the non-\CP , \cpp\
and \cpm samples for ({\it i}) the signal region, ({\it ii}) the $m_{D}$ sideband and ({\it iii}) the \DeltaE sideband.
In addition the signal region is divided into two samples according to the charge of the $B$ candidate.
We fit three probability density functions (PDF) weighted by the unknown event yields. 
For the \DeltaE sideband, we use $\mathcal{A}$. For the $m_{D}$ sideband~(sb) we use $a_{sb} \cdot
\mathcal{A}$ + $b_{sb} \cdot \mathcal{G}$, where $\mathcal{G}$ accounts for fake-$D$ candidates. 
For the signal region PDF, we use $a \cdot \mathcal{A}+ b \cdot \mathcal{G}_{peak}+ c \cdot \mathcal{G}_{signal}$,
where $b$  is scaled from $b_{sb}$ with the assumption that the number of fake $D$ background events present in 
the signal region is equal to the number measured in the $m_{D}$ sideband scaled by  the ratio of the 
$m_{D}$ signal-window to sideband widths, and $c$ is the number of $\Bpm\to D\Kstarpm$ signal events. The non-\CP mode 
sample, with relatively high statistics, helps constrain the PDF shapes for the low statistics \CP mode distributions. 
The \DeltaE sideband sample helps determine the $\mathcal{A}$ background shape. In total, the fit determines 19 event yields
as well as the mean and width of the signal Gaussian and the ARGUS parameter $\xi$.

Since the values of $\xi$ obtained for each data sample are consistent with each other, albeit with 
large statistical uncertainties, we have constrained $\xi$ to have the same value for all data samples in the fit. 
The simulation shows that the use of the same Gaussian parameters for all signal modes introduces only negligible 
systematic corrections. We assume that the fake $D$ backgrounds found in the $m_{D}$ sideband have the 
same final states as the signal and we fit these contributions with the same Gaussian parameterization.

\begin{figure}[t!] % Fit figures
  \begin{center}
    \epsfig{file=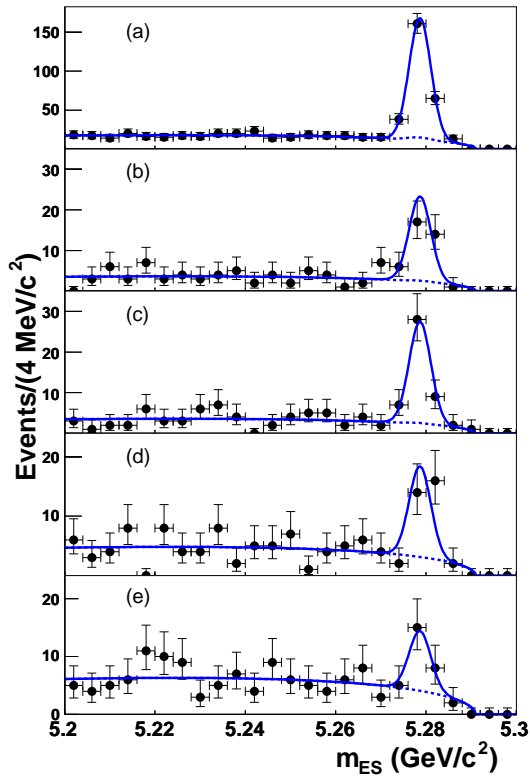,width=7.5cm}
    \caption{Distributions of \mes in the signal region for
      (a) the non-\CP modes in \Bpm decays,
      (b) the \cpp modes in \Bp, and (c) \Bm decays and  
      (d) the \cpm modes in \Bp and (e) \Bm decays. 
      The dashed curve indicates the total background contributions, which include 
      the fake $D$ backgrounds.
}
    \label{fig:asymmetryfit}
  \end{center}
\end{figure}

The fake $D$ background is assumed to not violate \CP and is therefore split equally between the \Bm and \Bp\ 
sub-samples. This assumption is consistent with results from our simulations and is considered further 
when we discuss the systematic uncertainties. The fit results are 
shown graphically in Fig.~\ref{fig:asymmetryfit} and numerically in Table~\ref{tab:nominalfitresult}. 
Table~\ref{tab:indivfitresult} records the number of events measured for each individual $D$ decay mode.

\begin{table}[htbp] % Fit table
  \begin{center}
    \caption{
      \label{tab:nominalfitresult}
      Results from the fit. For each GLW $D$ mode, we give the number of measured signal 
      events, the fake $D$ contribution, \Acp and \Rcp.  The fake $D$ contribution is calculated by scaling
      the number of fake $D$ events found in the $m_D$ sideband region to the signal region.
      The uncertainties are statistical only. We also show the number of measured signal events split by 
      the $B$ charge for $CP+$ and $CP-$ modes.}
    \begin{tabular}{ l c c c c} \hline \hline
                & \# Signal        & \# Fake~$D$ & \Acp                    & \Rcp                     \\ \hline 
      Non-\CP   &$231 \pm 17$      & 5.0              &                         &                          \\ \hline
      \cpp      &$\ 68.6 \pm\ 9.2$ & 0.3              &$~~\acppc \pm \acppstat$ & $~~\rcppc \pm \rcppstat$ \\
      ~~($B^+$) &$\ 31.2 \pm\ 6.2$ &                  &                         &                          \\
      ~~($B^-$) &$\ 37.4 \pm\ 6.8$ &                  &                         &                          \\ \hline
      \cpm      &$\ 38.5 \pm\ 7.0$ & 0.0              &$\acpmc \pm \acpmstat$   & $~~\rcpmc \pm \rcpmstat$ \\
      ~~($B^+$) &$\ 23.0 \pm\ 4.8$ &                  &                         &                          \\
      ~~($B^-$) &$\ 15.5 \pm\ 5.2$ &                  &                         &                          \\ \hline\hline
    \end{tabular}
  \end{center}
\end{table}
\begin{table}[htbp] % Individual fit table
  \begin{center}
    \caption{
      \label{tab:indivfitresult}
      Number of signal events from the GLW fit for individual $D$ decay modes studied in this analysis. We also 
      provide the selection efficiencies (in \%) and total correction factors ($\epsilon_c$). 
The uncertainties are statistical only.}
    \begin{tabular}{ l c c c c} \hline \hline
                     & \# Signal    &~~& Selection Efficiency (\%)&$\epsilon_{c}~(10^{-4})$ \\ \hline 
      Non-\CP        &              &&                      \\
      ~~$K^-\pip$    & 231 $\pm$ 17 &~~&12.76 $\pm$ 0.09&48.48 $\pm$ 0.96        \\ \hline
      \cpp           &              &                      \\
      ~~$K^+K^-$     & ~41 $\pm$ ~7 &~~&12.78 $\pm$ 0.05&4.90 $\pm$ 0.16       \\
      ~~$\pip\pim$   & ~28 $\pm$ ~6 &~~&12.34 $\pm$ 0.05&1.72 $\pm$ 0.11        \\ \hline
      \cpm           &              &&                      \\
      ~~$\KS\piz$    & ~21 $\pm$ ~7 &~~&~5.59 $\pm$ 0.03&4.10 $\pm$ 0.50        \\
      ~~$\KS\phi$    & ~~8 $\pm$ ~3 &~~& ~8.90 $\pm$ 0.04&1.30 $\pm$ 0.11        \\
      ~~$\KS\omega$  & ~~9 $\pm$ ~4 &~~&~2.35 $\pm$ 0.02&1.49 $\pm$ 0.30        \\ \hline\hline
    \end{tabular}
  \end{center}
\end{table}

Although most systematic uncertainties cancel for \Acp, a charge asymmetry inherent to the detector or data processing may 
exist. We quote the results from the study carried out in Ref.~\cite{detAsym}, where we used $B^- \ra \Dz\pi^-$ 
(with $\Dz$ decays into $CP$ or non-$CP$ eigenstates) events from control samples of data and simulation to 
measure the charge asymmetry. An average charge asymmetry of $A_{ch}=(-1.6\pm0.6)$\% was measured. We add linearly 
the central value and one-standard deviation in the most conservative direction to assign a systematic uncertainty of 
0.022. The second substantial systematic effect is a possible \CP~asymmetry in the fake $D$ background 
that cannot be excluded due to $CP$ violation in charmless $B$ decays. If there is an asymmetry 
$\mathcal{A}_{{\rm fake}~D}$, then the systematic uncertainty on \Acp\ is $\mathcal{A}_{{\rm fake}~D} \times b/c$, 
where $b$ is the contribution of the fake $D$ background and $c$ the signal yield. Assuming conservatively that 
$|\mathcal{A}_{{\rm fake}~D}| \leq 0.5$, we obtain systematic uncertainties of $\pm 0.003$ and $\pm 0.040$ on \Acpp\ and 
\Acpm respectively. Note that since we do not observe any fake $D$ background in $CP-$ modes, we use the 
statistical uncertainty of the signal yield from the fit to estimate this systematic uncertainty.

Since \Rcp\ is a ratio of rates of processes with different final states of the $D$, we must consider the 
uncertainties affecting the selection algorithms for the different $D$ channels. This results in small 
corrections which account for the difference between the actual detector response and the simulation 
model. The main effects stem from the approximate modeling of the tracking efficiency (a correction of 0.4\% 
per pion track coming from a \KS and 0.2\% per kaon and pion track coming from other candidates), the \KS 
reconstruction efficiency for \cpm modes of the \Dz\ (1.3\% per \KS in $\KS\phi$ mode and 2.0\% in $\KS\piz$ 
and $\KS\omega$), the \piz reconstruction efficiency for the $\KS\piz$ and $\KS[\pip\pim\piz]_{\omega}$ channels 
(3\%) and the efficiency and misidentification probabilities from the particle identification (2\% per track). 
The corrections  are calculated by comparing data and Monte Carlo using high-statistics and high-purity 
samples. Charged kaon and pion samples obtained from $D$-meson decays ($D^{*+} \ra \Dz \pip$) are used for 
particle identification corrections. For tracking corrections, we use $\tau$-pair events where one $\tau$ decays 
to a muon and two neutrinos and the other decays to $\rho^0 h \nu$ where $h$ is a $K$ or a $\pi$. $\Bz \ra \phi \KS$ and 
$\Bz \ra \pi^+D^- (D^- \ra \KS\pi^-)$ decays are used for \KS corrections, and \piz correction factors are 
calculated using $\tau \ra \rho \nu$ and $\tau \ra \pi \nu$ samples. 
The total correction factors, which also include branching fractions and selection efficiencies, 
used in the calculation of \Rcppm (Eq.~\ref{equation:Rcal}) are given in Table~\ref{tab:indivfitresult}.
$R_{CP+}$ ($R_{CP-}$) is calculated using the sum of the individual $CP+$ ($CP-$) correction factors.
Altogether, the systematic uncertainties 
due to the corrections equal $\pm0.078$ and $\pm0.100$ for \Rcpp\ and \Rcpm, respectively. 
The uncertainties on the measured branching fractions~\cite{pdg} and efficiencies for different $D$ 
decay modes, are included in the calculation of the systematic errors due to these corrections.

Another systematic correction applied to the \cpm measurements arises from a possible \cpp 
background in the $\KS\phi$ and $\KS\omega$ channels. In this case, the observed quantities 
${\cal A}^{{\rm obs}}_{\cpm}$ and ${\cal R}^{{\rm obs}}_{\cpm}$ are corrected:
\begin{eqnarray}
  \Acpm = ( 1 + \epsilon ) {\cal A}^{{\rm obs}}_{\cpm} - \epsilon  \Acpp ; \label{eq:Acorr} \ 
  \Rcpm =  \frac{{\cal R}^{{\rm obs}}_{\cpm} }{ ( 1 + \epsilon ) }, \label{eq:Rcorr} \nonumber
\end{eqnarray}\noindent
where $\epsilon$ is the ratio of \cpp background to \cpm signal. An investigation of the 
$\Dz\to\Km\Kp\KS$ Dalitz plot~\cite{KKKS} indicates that the dominant background for $\Dz\to\KS\phi$ comes 
from the decay $a_0(980)\to\Kp\Km$, at the level of $(25\pm 1)\%$ of the size of the $\phi\KS$ signal. We have no 
information for the $\omega\KS$ channel and assume $(30\pm30)\%$ of \cpp background contamination. The 
$\KS\piz$ mode has no $CP+$ background. 
The value of $\epsilon$ for the combination of $CP-$ modes is $(11\pm7)$\%. The systematic 
uncertainty associated with this effect is $\pm0.02$ and $\pm0.06$ for \Acpm and \Rcpm, respectively.

To account for the non resonant $\KS\pi^-$ pairs in the $K^*$ mass range we study a model that incorporates 
$S$-wave and $P$-wave pairs in both the $b\to c\overline{u}s$ and $b\to u\overline{c}s$ amplitudes. The $P$-wave mass 
dependence is described by a single relativistic Breit-Wigner while the $S$-wave component is assumed to be a 
complex constant. It is expected that higher order partial waves will not contribute significantly and therefore 
they are neglected in the model. We also assume that the same relative amount of $S$ and $P$-wave is present in the
 $b\to c\overline{u}s$ and $b\to u\overline{c}s$ amplitudes.
The amount of $S$-wave present in the favored $b\to c\overline{u}s$ amplitude is 
determined directly from the data by fitting the angular distribution of the $\KS\pi$ system in the 
$\Kstar$ mass region, accounting for interference~\cite{DKstarBF}. From this fit we determine that the number of  non-$K^*$ $K_S\pi^-$ events is 
(4 $\pm$ 1)\% of the measured signal events.
 To estimate the systematic uncertainties due to this source we vary all the strong 
phases between 0 and 2$\pi$ and calculate the maximum deviation between the $S$-wave model and the expectation if 
there were no non-resonant contribution for both \Acppm [Eq.~(\ref{eq:A})] and \Rcppm [Eq.~(\ref{eq:R})]. 
This background induces systematic variations of $\pm 0.051$ for \Acppm and $\pm 0.035$ for \Rcppm. 

The last systematic uncertainty is due to the assumption that the  parameters of the Gaussian and ARGUS functions are the same throughout 
the signal region, $\Delta E$ and $m_{D}$ sidebands. We estimate the uncertainties by varying the width and mean of the 
Gaussian and $\xi$ of the ARGUS by their corresponding statistical uncertainties obtained from the fit. All the 
systematic uncertainties are listed in Table \ref{tab:GLW}. We add them in quadrature and quote the final results:
\begin{flalign*}
  \Acpp &= ~~\acppc  \pm \acppstat(\mathrm{stat.})  \pm \acppsys(\mathrm{syst.}) \\
  \Acpm &= \acpmc  \pm \acpmstat(\mathrm{stat.})  \pm \acpmsys(\mathrm{syst.}) \\
  \Rcpp &= ~~\rcppc \pm \rcppstat(\mathrm{stat.})  \pm \rcppsys(\mathrm{syst.}) \\
  \Rcpm &= ~~\rcpmc  \pm \rcpmstat(\mathrm{stat.}) \pm \rcpmsys(\mathrm{syst.})
\end{flalign*}
\begin{table}[h]
  \begin{center}
    \caption{\label{tab:GLW}Summary of systematic uncertainties for the GLW analysis.}
    \begin{tabular}{lcccc} \hline \hline
      Source                                     & $\delta \Acpp$ & $\delta \Acpm$ & $\delta \Rcpp$ & $\delta \Rcpm$ \\ \hline
      Detection asymmetry                        & $0.022$     & $0.022$     & -              & -              \\
      Non-resonant $K_s^0\pi^-$ bkg.             & $0.051$     & $0.051$     & $0.035$     & $0.035$     \\
      Same-final-state bkg.                      & -              & $0.019$     & -              & $0.061$     \\
      Asymmetry in fake $\Dz$ bkg.               & $0.003$     & $0.040$     & -              & -              \\
      Efficiency correction                      & -              & -              & $0.078$     & $0.100$     \\
      Same $\mathcal{G}$ and $\mathcal{A}$ shape & $0.003$     & $0.013$     & $0.009$     & $0.025$     \\ \hline
      Total systematic uncertainty                     & $0.056$     & $0.072$     & $0.086$     & $0.125$     \\
      \hline \hline
    \end{tabular}
  \end{center}
\end{table}

These results can also be expressed in terms of $x_\pm$ defined in Eq.~(\ref{xvariables}):

\begin{eqnarray}
  x_+ &=& 0.21 \pm 0.14 {\rm (stat.)} \pm 0.05 {\rm (syst.)}, \nonumber \\
  x_- &=& 0.40 \pm 0.14 {\rm (stat.)} \pm 0.05 {\rm (syst.)}, \nonumber
\end{eqnarray}

\noindent
where the \cpp pollution systematic effects are included. Including these effects increased $x_+$ and $x_-$ by  
$0.035 \pm 0.024$ and $0.023 \pm 0.017$, respectively.

%*******************************
\section{\boldmath The ADS Analysis}
\begin{figure}[ht]
  \begin{center}
    \includegraphics[width=9cm]{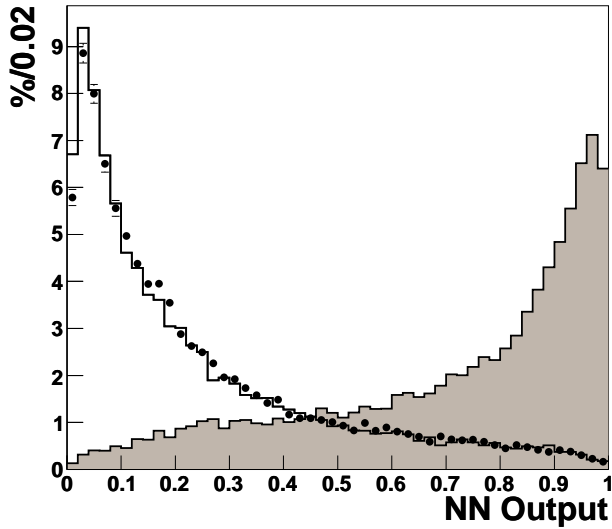}
    \caption{\label{fig:ADSnnPerformance}
      Neural network (NN) output and result of the NN verification for 
      the ADS analysis (see text). 
      The samples used to produce the output are shown as histograms. 
      The signal (Monte Carlo simulation) is the shaded histogram peaking 
      near 1; the continuum (Monte Carlo simulation) is the histogram peaking near 0.
      The off-peak data 
%recorded at $\sqrt{s}$ 40 \mev below the resonance 
used to check the  NN are overlaid as data points.}
  \end{center}
\end{figure}

%*******************************
In the ADS analysis we only use $D$ decays with a charged kaon and pion in the final state
and $K^{*-}$  decays to $\KS\pi^-$ followed by $\KS \to \pi^+\pi^-$. 
The ADS event selection criteria and procedures are nearly identical to those used for the GLW analysis. 
However due to the small value of $r_D$ the yield of interesting ADS events (i.e. $B^-\to [K^+\pi^-]_D K^{*-}$ 
and  $B^+\to [K^-\pi^+]_D K^{*+}$) is expected to be smaller than for the GLW analysis. Therefore in 
order to reduce the background in the ADS analysis the $K^0_S$ invariant mass window is narrowed
to 10 $\mevcc$ and the $K^{*-}$ invariant mass cut is reduced to 55 $\mevcc$. A neural network using the
same variables as in the GLW analysis is trained on ADS signal and continuum MC events and verified 
using off-peak continuum data. The separation between signal and continuum background is shown in 
Fig.~\ref{fig:ADSnnPerformance}. We select candidates, both right and wrong-sign,
 with neural network output above 0.85. All other 
$K^0_S$, $K^{*-}$, and continuum suppression criteria are the same as those used in the GLW analysis.

 $D\to\Km\pip$ and $\Kp \pim$ candidates are used in this analysis. Candidates that have 
an invariant mass within 18 $\mevcc$ (2.5 standard deviations) of the nominal $D^0$ mass~\cite{pdg} are 
kept for further study. We require kaon candidates to pass the same particle identification criteria 
as imposed in the GLW analysis.

We identify $B$-meson candidates using the beam-energy-substituted mass $\mes$ and the energy 
difference $\Delta E$. For this analysis signal candidates must satisfy $|\Delta E|\le 25$ \mev. 
The efficiency to detect a $\Bm \to \Dz \Kstarm$ signal event where 
$\Dz \to K \pi$, after all criteria are imposed, is (9.6$\pm$0.1)\% and is 
the same for $D^0\to K^-\pi^+$ and $D^0\to K^+\pi^-$.  
In 1.8\% of the events we find more than one suitable candidate. In such cases we choose the candidate with the 
smallest $\chi^2$ defined in Eq.~(\ref{equation:multiChi2}). Simulations show that negligible bias is introduced by this 
choice and the correct candidate is picked about 88\% of the time.

We study various potential sources of background using a combination of Monte Carlo simulation 
and data events. Two sources of background are identified in large samples of simulated \BB 
events. One source is $B^- \ra \Dz\KS\pim$ production where the $\KS\pi^-$ is nonresonant and has an 
invariant mass in the $K^{*-}$ mass window. This background is discussed later. The 
second background (peaking background) includes instances where a favored decay (e.~g.~ 
$B^-\to [K^-\pi^+]_DK^{*-}$) contributes to fake candidates for the suppressed decay (i.e. 
$B^+\to [K^-\pi^+]_DK^{*+}$). The most common way for this to occur is for a $\pi^+$ from the rest 
of the event to be substituted for the $\pi^-$ in the $K^{*-}$ candidate. Other sources of peaking 
background include double particle-identification failure in signal events that results in 
$D^0\to K^-\pi^+$ being reconstructed as $D^0\to \pi^-  K^+$, or the kaon from the $D^0$ being 
interchanged with the charged pion from the $K^*$. We quantify this background with the ratio of the signal 
efficiency of wrong-sign decay to right-sign decay multiplied by the right-sign yield from 
data. The total size of this right-sign pollution is estimated to be $2.4~\pm~0.3$ events. 
Another class of backgrounds is
charmless decays with the same final state as the signal (e.g., $B^-\to K^{*-}K^+\pi^-$). 
Since the branching fractions for many of these charmless decays 
have not been measured 
or are poorly measured, we use the $D$ sideband to estimate the contamination from 
this source. From a fit to the $\mes$ distribution using candidates in the $D$ sideband we find $0.0\pm 1.1$ 
events. We take the 1.1 events as the contribution to the systematic uncertainty from this source.

Signal yields are determined from an unbinned extended maximum likelihood fit to the $\mes$ distribution
in the range $5.2 \le \mes \le 5.3\ \gevcc$. A Gaussian function ($\mathcal{G}$) is used to describe 
all signal shapes while the combinatorial background is modeled with an ARGUS threshold function 
($\mathcal{A}$) defined in Eq.~(\ref{equation:ARGUS}). The mean and width of the Gaussian as well
as the $\xi$ parameter of the ARGUS function are determined by the fit. For the likelihood function we use 
$a\cdot \mathcal{A}+b\cdot \mathcal{G}$ where $a$ is the number of background events and $b$ the number of 
signal events. We correct $b$ for the right-sign peaking background previously discussed (2.4$\pm$0.3 events).
\begin{figure}[htb]
  \begin{center}
    \includegraphics[width=6cm]{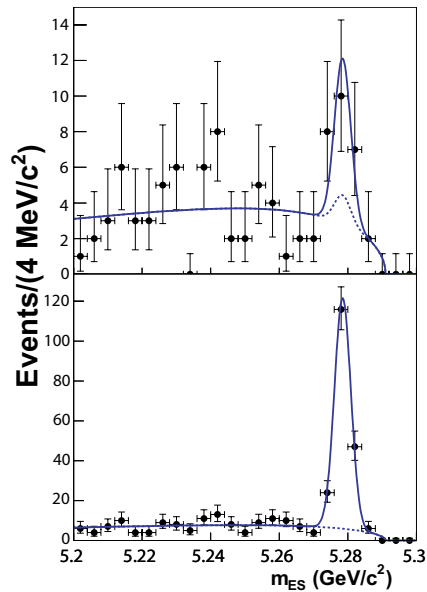}
    \caption{\label{fig:adsfit}
      Distributions of \mes for the wrong-sign (top) and right-sign 
      (bottom) decays. These decay categories are defined in the text. 
      The dashed curve indicates the total background contribution. It also includes 
      the right-sign peaking background estimated from a Monte Carlo study 
      for the wrong-sign (top) decays. The curves result from a simultaneous fit to 
      these distributions with identical PDFs for both samples.}
  \end{center}
\end{figure}
\begin{figure}[hbt]
  \begin{center}
    \includegraphics[width=6cm]{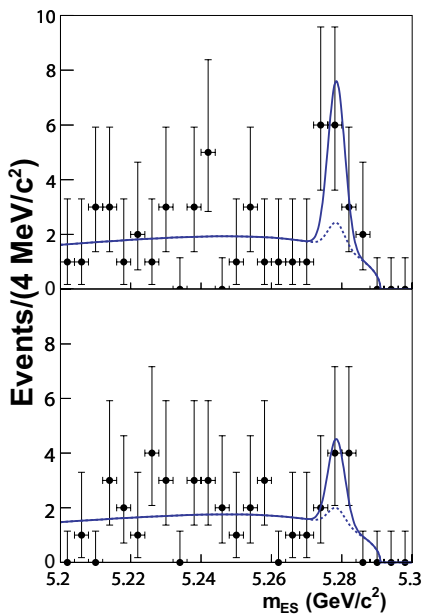}
    \caption{\label{fig:adsfitA}
      The  wrong-sign sample shown 
      in the top plot of Fig.~\ref{fig:adsfit} split by charge.
      Upper plot shows the \mes distribution of the $B^+\to [K^-\pi^+]_DK^{*+}$ decays 
      while  lower plot presents the same for the $B^-\to [K^+\pi^-]_DK^{*-}$ 
      decays. The dashed curve indicates the total background contribution which includes 
      the right-sign peaking background. 
}
  \end{center}
\end{figure}

In Fig.~\ref{fig:adsfit} we show the results of a simultaneous fit to $B^-\to [K^+\pi^-]_DK^{*-}$ and
$B^-\to [K^-\pi^+]_DK^{*-}$ candidates that satisfy all selection criteria. 
It is in the wrong-sign 
decays that the interference we study takes place.  Therefore in Fig.~\ref{fig:adsfitA} we display the 
same fit separately for the wrong-sign decays of the \Bp and the \Bm mesons. The results of the maximum 
likelihood fit are $\Rads =0.066\pm0.031$, $\Aads =-0.34\pm0.43$, and $172.9 \pm 14.5 \ B^-\to [K^-\pi^+]_DK^{*-}$ 
right-sign events. Expressed in terms of the wrong-sign yield, the fit result is $11.5 \pm 5.3$ wrong-sign 
events (3.8 $\pm$ 3.4 $B^- \ra [K^+\pim]_D K^{*-}$ and 7.7 $\pm$ 4.2 $B^+ \ra [K^-\pip]_D K^{*+}$ events). 
The uncertainties are statistical only. The correlation between \Rads and \Aads is insignificant. 

We summarize in Table~\ref{tab:ADS} the systematic uncertainties relevant to this analysis. Since both 
\Rads and \Aads are ratios of similar quantities, most potential sources of systematic uncertainties cancel.

For the estimation of the detection-efficiency asymmetry we use the previously mentioned results from the study 
carried out in Ref.~\cite{detAsym}. 
We add linearly the central value and one-standard deviation in 
the most conservative direction to assign a systematic uncertainty of $\delta A_{ch}=\pm0.022$ to the \Aads 
measurement. To a good approximation the systematic uncertainty in \Rads due to this source is 
$\delta \Rads = \Rads \cdot \Aads \cdot \delta A_{ch}$.

To estimate the systematic uncertainty on \Aads and \Rads due to the peaking background, we use the 
statistical uncertainty on this quantity, $\pm0.3$ events. With approximately $12\ B^-\to [K^+\pi^-]_DK^{*-}$ 
events and $173\ B^-\to [K^-\pi^+]_DK^{*-}$ events this source contributes $\pm0.002$ and 
$\pm0.024$ to the systematic uncertainties on \Rads and \Aads, respectively.
Similarly, the 1.1 events uncertainty on the same-final-state background leads to
systematic uncertainties of $\pm 0.0061$ and $\pm 0.091$ on \Rads and \Aads respectively.

As in Section~\ref{GLW}, we need to estimate the systematic effect due to the non-resonant $\KS\pi^-$ pairs in 
the $K^*$ mass range. We follow the same procedure discussed in Section~\ref{GLW}. After adding in quadrature 
the individual systematic uncertainty contributions, listed in Table~\ref{tab:ADS}, we find:
\begin{flalign*}
  \Aads &= \aadsc \pm \aadsstat(\mathrm{stat.}) \pm \aadssys(\mathrm{syst.}) \\
  \Rads &= ~~\radsc \pm \radsstat(\mathrm{stat.}) \pm \radssys(\mathrm{syst.}).
\end{flalign*}
\begin{table}[h]
  \begin{center}
    \caption{\label{tab:ADS}Summary of ADS systematic uncertainties.}
    \begin{tabular}{lcc} \hline \hline
      Source                         & $\delta \Rads$ & $\delta \Aads$ \\ \hline
      Detection asymmetry            & $\pm0.0005$    & $\pm0.022$ \\
      Peaking bkg.                   & $\pm0.0020$    & $\pm0.024$ \\
      Same-final-state bkg.          & $\pm0.0061$    & $\pm0.091$ \\
      Non resonant $K_s^0\pi^-$ bkg. & $\pm0.0073$    & $\pm0.126$ \\ \hline
      Total systematic uncertainty   & $\pm0.0097$    & $\pm0.159$ \\ \hline \hline
    \end{tabular}
  \end{center}
\end{table}
%*******************************
\section{Combined Results}
\begin{figure}[hbt]
  \begin{center}
   \includegraphics[width=9cm]{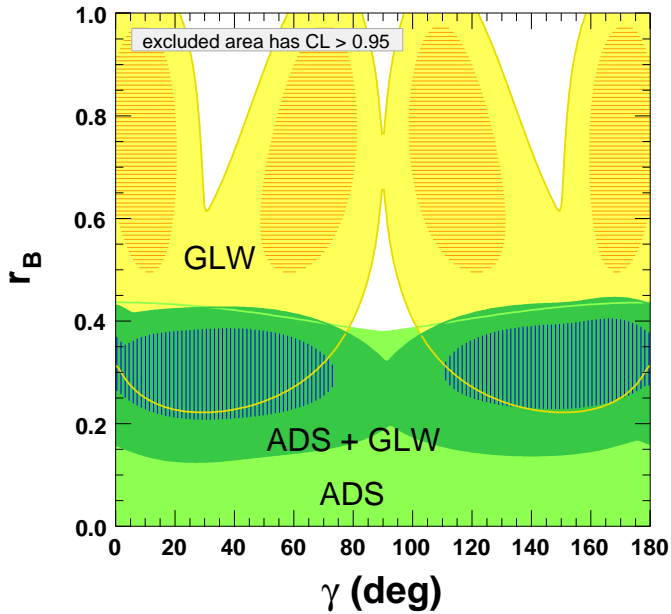}
   \caption{\label{fig:rbVSgamma}
95\% confidence level contours from a two dimensional scan of $\gamma$ versus $r_B$ from the $\Bm\to DK^{*-}$ GLW and ADS measurements.
Also shown are the 68\% confidence level regions (striped areas) for the GLW and the fit which uses
both the GLW and ADS measurements. $r_D$ and $\delta_D$ are from Ref.~\cite{HFAG2008}.
}
  \end{center}
\end{figure}

%*******************************
We use the GLW and ADS results and a frequentist statistical 
approach~\cite{CKMfitter} to extract information on $r_B$ and $\gamma$.
In this technique, a $\chi^2$ is calculated using the differences between the
measured and theoretical values and the statistical and systematic errors
 of the six measured
quantities. The values of $r_D$ and $\delta_D$ are taken from Ref.~\cite{HFAG2008},
while we allow $0\leq r_B\leq 1$, $0^{\circ}\leq \gamma \leq 180^{\circ}$, and 
$0^{\circ}\leq \delta_B \leq 360^{\circ}$.   
The minimum of the $\chi^2$ for the $r_B$, $\gamma$, and $\delta_B$ parameter space is calculated first ($\chi^2_{{\rm min}}$).
We then scan the range of $r_B$ and $\gamma$ minimizing the $\chi^2$ ($\chi^2_{{\rm m}}$) by varying $\delta_B$.
A confidence level for each value of $r_B$ and $\gamma$ is  calculated using $\Delta \chi^2=\chi^2_{\rm m}-\chi^2_{\rm min}$ and one degree
of freedom. We
assume Gaussian measurement uncertainties
and confirm this assumption using simulations. 
In Fig.~\ref{fig:rbVSgamma}
we show the 95\% confidence level contours for  $r_B$ versus $\gamma$ as well as the 68\% confidence level
contours for the GLW and the combined GLW and ADS analysis (striped areas).   

\begin{figure}[ht]
  \begin{center}
    \includegraphics[width=9cm]{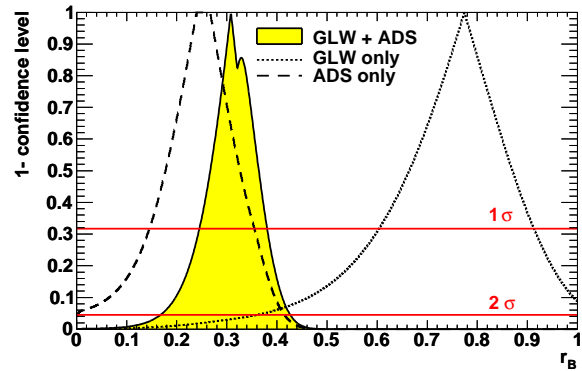} 
    \caption{\label{fig:rB}
      Constraints on $r_B$ from the $\Bm\to DK^{*-}$ GLW and ADS measurements. 
      The dashed (dotted) curve shows  1 minus the confidence level to exclude the abscissa value as a 
function of $r_B$ derived from the GLW (ADS) measurements. The GLW+ADS result (solid line and shaded area) is from a fit which uses
both the GLW and ADS measurements as well as $r_D$ and $\delta_D$ from~\cite{HFAG2008}. 
The horizontal lines show the exclusion limits at the  1, and 2 standard deviation levels. }
  \end{center}
\end{figure}

In order to find confidence levels for $r_B$ we use the above procedure, 
minimizing $\chi^2$ with respect to $\gamma$ and $\delta_B$.
 The results of this calculation are shown in Fig.~\ref{fig:rB}.
The fit which uses both the ADS and GLW results has its minimum  $\chi^2$ at $r_B=0.31$ with a one sigma interval of [0.24, 0.38]
and a two sigma interval of [0.17, 0.43]. The value $r_B=0$ is excluded at the 3.3 sigma level.
 We find similar results
for $r_B$ using the modifications to this frequentist approach discussed in Ref.~\cite{GGZ} and using the Bayesian approach of
Ref.~\cite{UTFIT}. 

\begin{figure}[hbt]
  \begin{center}
   \includegraphics[width=9cm]{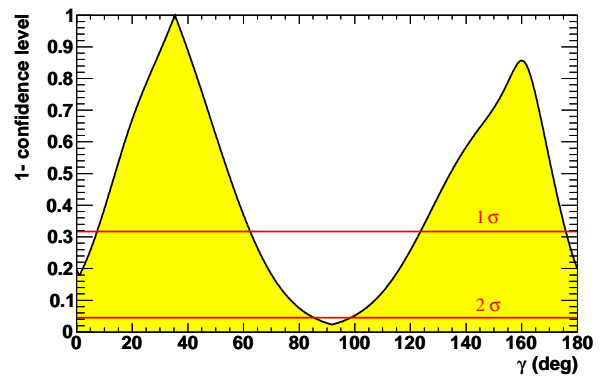} 
   \caption{\label{fig:gamma}
Constraints on $\gamma$ from a fit which uses
both the $\Bm\to DK^{*-}$  GLW and ADS measurements as well as $r_D$ and $\delta_D$ from~\cite{HFAG2008}.  
The horizontal lines show the exclusion limits at the  1 and 2 standard deviation levels.     
}
  \end{center}
\end{figure}

Using the above procedure we also find confidence intervals for $\gamma$. The results of the scan in $\gamma$ are shown in
Fig.~\ref{fig:gamma}. The combined GLW+ADS analysis excludes
values of $\gamma$ in the regions $[0,~7]^{\circ},~[62,~124]^{\circ}$ and $[175,~180]^{\circ}$ at the one sigma
 level and $[85,~99]^{\circ}$ at the two sigma level. 
The use of the 
measurement of the strong
phase $\delta_D$~\cite{HFAG2008} helps to resolve the ambiguities on $\gamma$ 
and therefore
explains the asymmetry in the confidence level plot shown in Fig.~\ref{fig:gamma}.

%%%%%%%%%

%

%*******************************
\section{Summary}
%*******************************
In summary, we present improved  measurements of yields from $B^-\ra D K^{*-}$ decays, where 
the neutral $D$-meson decays into final states of even and odd \CP (GLW), and the $K^+\pi^-$ final state (ADS). 
We express the results as \Rcp, \Acp, $x_\pm$, \Rads, and \Aads. The value $r_B$ = 0  is excluded 
at the 3.3 sigma level. These results in combination with other 
GLW, ADS, and Dalitz type analyses  improve our knowledge of $r_B$ and $\gamma$.
\section{Acknowledgments}
We are grateful for the 
extraordinary contributions of our \pep2\ colleagues in
achieving the excellent luminosity and machine conditions
that have made this work possible.
The success of this project also relies critically on the 
expertise and dedication of the computing organizations that 
support \babar.
The collaborating institutions wish to thank 
SLAC for its support and the kind hospitality extended to them. 
This work is supported by the
US Department of Energy
and National Science Foundation, the
Natural Sciences and Engineering Research Council (Canada),
the Commissariat \`a l'Energie Atomique and
Institut National de Physique Nucl\'eaire et de Physique des Particules
(France), the
Bundesministerium f\"ur Bildung und Forschung and
Deutsche Forschungsgemeinschaft
(Germany), the
Istituto Nazionale di Fisica Nucleare (Italy),
the Foundation for Fundamental Research on Matter (The Netherlands),
the Research Council of Norway, the
Ministry of Education and Science of the Russian Federation, 
Ministerio de Educaci\'on y Ciencia (Spain), and the
Science and Technology Facilities Council (United Kingdom).
Individuals have received support from 
the Marie-Curie IEF program (European Union) and
the A. P. Sloan Foundation.

%%%%%%%%%%%%%%%%%%%%%%%%%%%%%%%%%%%%%%%%%%%%%%%%%%%%%%%%   


\begin{thebibliography}{99}
\bibitem{892}
Hereafter $K^{*-}$ implies $K^*(892)^-$. In this paper we use only the  $K^{*\pm}\to\KS\pi^{\pm}$, $\KS\to \pi^+\pi^-$ 
decay chain.
\bibitem{D0D0bar}
By $D$ we mean any linear combination of $\Dz$ and  $\Dzb$.
\end{thebibliography}
\end{document}